# MR-Based PET Attenuation Correction using a Combined Ultrashort Echo Time/Multi-Echo Dixon Acquisition

Paul Kyu Han[1,2,†], Debra E. Horng[1,2,†], Kuang Gong[1,2], Yoann Petibon[1,2], Kyungsang Kim[1,2], Quanzheng Li[1,2], Keith A. Johnson[1,2,3,4], Georges El Fakhri[1,2], Jinsong Ouyang[1,2], and Chao Ma[1,2*]

[1]Gordon Center for Medical Imaging, Department of Radiology, Massachusetts General Hospital, Boston, MA 02114

[2]Department of Radiology, Harvard Medical School, Boston, MA 02115

[3]Department of Neurology, Massachusetts General Hospital, Boston, MA 02114

[4]Center for Alzheimer Research and Treatment, Department of Neurology, Brigham and Women's Hospital, Boston, MA 02115

Running Title: MR-Based PET AC using mUTE

*Correspondence to:    Chao Ma, PhD
                       Gordon Center for Medical Imaging
                       Massachusetts General Hospital
                       125 Nashua Street, Suite 660
                       Boston, MA, 02114
                       Telephone: +1-617-643-1961
                       E-mail: cma5@mgh.harvard.edu

†These authors contributed equally to this work.

# Abstract

**Purpose:** To develop a magnetic resonance (MR)-based method for estimation of continuous linear attenuation coefficients (LAC) in positron emission tomography (PET) using a physical compartmental model and ultrashort echo time (UTE)/multi-echo Dixon (mUTE) acquisitions.

**Methods:** We propose a three-dimensional (3D) mUTE sequence to acquire signals from water, fat, and short-$T_2$ components (e.g., bones) simultaneously in a single acquisition. The proposed mUTE sequence integrates 3D UTE with multi-echo Dixon acquisitions and uses sparse radial trajectories to accelerate imaging speed. Errors in the radial k-space trajectories are measured using a special k-space trajectory mapping sequence and corrected for image reconstruction. A physical compartmental model is used to fit the measured multi-echo MR signals to obtain fractions of water, fat and bone components for each voxel, which are then used to estimate the continuous LAC map for PET attenuation correction.

**Results:** The performance of the proposed method was evaluated via phantom and in vivo human studies, using LACs from Computed Tomography (CT) as reference. Compared to Dixon- and atlas-based MRAC methods, the proposed method yielded PET images with higher correlation and similarity in relation to the reference. The relative absolute errors of PET activity values reconstructed by the proposed method were below 5% in all of the four lobes (frontal, temporal, parietal, occipital), cerebellum, whole white matter and gray matter regions across all subjects (n=6).

**Conclusions:** The proposed mUTE method can generate subject-specific, continuous LAC map for PET attenuation correction in PET/MR.

# Introduction

Position emission tomography (PET) / magnetic resonance imaging (MRI) is an emerging imaging modality that shows great potential in clinical applications. PET/MR allows high-resolution anatomical imaging with excellent soft-tissue contrast as well as functional and molecular information. PET/MR also provides unique opportunities for improving the quality of PET imaging via incorporating motion field and anatomical prior information from MR into PET image reconstruction[1-3]. Furthermore, combining PET and MR in a single imaging session enables estimation of physiological processes that otherwise would be impossible with PET or MR alone[4].

Despite the usefulness and growing interest of PET/MR imaging, PET attenuation correction with MR is not a trivial task. MR signal is a complex function of many variables such as proton density and relaxation times but not the electron density, which determines the linear attenuation coefficients (LAC) of different tissues. This fundamental lack of information makes PET attenuation correction with MR difficult. The issue becomes obvious in tissues such as bone, which displays high LAC due to the high electron density but low MR signal due to the very short $T_2$ decay and low proton density.

Various MR-based attenuation correction (MRAC) methods have been developed to address this issue[5,6]. Early developments used MR images to segment/classify voxels into different tissue classes and then assigned a specific LAC value to each class for PET attenuation correction. The MRAC method implemented in the first generation of commercial PET/MR scanners segments the imaging object into 3-tissue (i.e., air, soft-tissue, and lung)[7] or 4-tissue classes (i.e., air, fat, non-fat soft-tissue, and lung)[8,9] based on $T_1$-weighted or two-point Dixon MR images. Methods employing ultrashort echo time (UTE) or zero echo time (ZTE) acquisitions have also been developed to detect bone signal and assign a single LAC to bone tissue[10-14]. However, these early methods do not properly account for continuous variations in LAC and suffer from biases in the resultant PET reconstructions[15].

Several MRAC methods have been developed to estimate LAC maps with continuous distribution. One family of methods exploit a pre-compiled template (i.e., atlas) generated from Computed Tomography (CT)-MR databases and employ image registration to align acquired MR images with the template and get LAC maps with continuous distribution. A partial list of these methods include deformable registration methods[16,17] with evaluation of morphologic similarity[18], machine learning approaches such as Gaussian process regression[16,19], Gaussian mixture regression model[20], support vector regression model[21], structured random forest[22], and patch-based approaches exploiting local subregions of images rather than the whole image[23-26]. Recently, tissue segmentation information has also been proposed for improved estimation of LAC[27-31]. Although these atlas-based MRAC methods work well in general for brain imaging of subjects with normal anatomy, these methods are known to be sensitive to registration errors, cannot account for inter-subject variability, and cannot properly handle subject-specific anatomical abnormality that differs from the template.

Another family of methods estimate LAC maps with continuous distribution by modeling the relationship between MR and CT information. Methods employing ZTE acquisition have been developed to model the relationship between ZTE signal intensities and CT Hounsfield values to obtain pseudo-CT images with continuous distribution of bone[31-35]. Methods employing UTE acquisition have been developed to model the relationship between relaxation time (e.g., $R_2$*) information and CT Hounsfield values[36,37] or LAC[38] to obtain pseudo-CT images or LAC maps with continuous distribution. However, these methods rely on empirical relationship derived from MR and CT images, and require MR and CT dataset pairs to establish the relationship.

Recently, deep neural networks (DNN) have also been proposed for MRAC to generate LAC maps with continuous distribution. Algorithms based on feed forward neural network (FFNN)[39], general adversarial network (GAN)[40,41], and convolutional neural network (CNN) with convolutional auto encoder (CAE)[14,42] or U-net structure[43-46] with group convolution modules[47] have shown encouraging results in generating subject-specific LAC maps with continuous distribution. These methods generate pseudo-CT images directly from conventional MR images (e.g., Dixon or $T_1$-weighted images) which are later converted to LAC maps through bilinear transformation. Though promising, one major drawback of these methods is in the requirement of substantial number of MR and CT dataset pairs to train and generate LAC maps, which may not be easily available for specific applications, e.g., oncologic applications.

In this work, we propose a new physical model-based MRAC method to estimate subject-specific LAC maps with continuous distribution. More specifically, we propose a combined UTE/multi-echo Dixon (mUTE) sequence along with a physical compartmental model to estimate the different proportions of water, fat and bone components in each voxel, which are subsequently used to estimate LAC with continuous distribution. The proposed method (i) allows to estimate continuous LAC directly from MR through physical compartmental modeling, (ii) does not require any training datasets to model relationships or to generate subject-specific LAC maps, (iii) provides robust fat-water separation that does not require registration by performing UTE and multi-echo Dixon in a single acquisition, and (iv) is not mutually exclusive to other MRAC methods. The feasibility of the proposed method is demonstrated via a phantom experiment with pig tibia and in vivo experiments with 6 subjects (1 patient with mild-cognitive impairment and 5 cognitively healthy volunteers) undergoing brain MR and PET/CT examinations using [11]C-Pittsburgh compound B (PiB). Potential usefulness of the proposed method and limitations of the current work are also discussed.

## Materials and Methods

### A Physical Compartmental Model for Continuous LAC

The proposed work is based on a physical compartmental model to estimate LAC using MR. A single imaging voxel is assumed to consist of water, fat, bone and air compartments, with different proportions of each of the compartments. The LAC at a single voxel ($\mu_{voxel}$) can be defined as:

$$\mu_{voxel} = p_W \cdot \mu_W + p_F \cdot \mu_F + p_B \cdot \mu_B \qquad [1]$$

where $p_W$, $p_F$, and $p_B$ each denote the volume fractions of water, fat and bone, respectively, and $\mu_W$, $\mu_F$, and $\mu_B$ each denote the LAC of water, fat, and bone, respectively. Note that the LAC of air is negligible and therefore ignored in Eq. 1. In this model, water and fat refer to those not only within soft-tissue but also the long-$T_2$ components of water and fat within the bone, respectively. The advantage of this physical compartmental model is estimation of the continuous variations of LAC via the fraction of the compartments, which can be estimated with MR using the strategy proposed thereafter.

### Proposed 3D mUTE Sequence

Our proposed three-dimensional (3D) mUTE sequence combines 3D UTE[48] and multi-echo Dixon imaging[49,50] in a single acquisition (Fig. 1). Multi-echo radial images are acquired at different TE times every other repetition time (TR). The following signal model is used to estimate the proton densities of water, fat, and bone from the acquired multi-echo images:

$$S(TE_j) = \begin{cases} \left(\rho_B + \left(\rho_W + \rho_F \cdot g(TE_j)\right) \cdot e^{-R_2^* \cdot TE_j}\right) \cdot e^{i \cdot 2\pi \cdot \Delta f_{B0} \cdot TE_j} & for\ j = 1 \\ \left(\rho_W + \rho_F \cdot g(TE_j)\right) \cdot e^{-R_2^* \cdot TE_j} \cdot e^{i \cdot 2\pi \cdot \Delta f_{B0} \cdot TE_j} & for\ j = 2, \dots, J \end{cases}$$ [2]

where $S(TE_j)$ denotes the signal at a single imaging voxel acquired at the $j^{\text{th}}$ TE time ($TE_j$); $\rho_W$, $\rho_F$, and $\rho_B$ each denote the proton densities of water, fat, and bone, respectively; $g(t_j) = \sum_{m=1}^{M} \alpha_m \cdot e^{i \cdot 2\pi \cdot (\Delta f_{fat,m}) \cdot t_j}$ denotes the $M$-peak (i.e., $M = 6$) spectral model of fat with relative amplitude $\alpha_m$ (i.e., $\sum_{m=1}^{M} \alpha_m = 1$) and frequency offset $\Delta f_{fat,m}$, where $\alpha_m$ and $\Delta f_{fat,m}$ are known from the literature[51,52]; $R_2^* = 1/T_2^*$ denotes the single representative relaxation rate of both water and fat[53]; and $\Delta f_{B0}$ denotes the frequency offset due to magnetic field inhomogeneity ($B_0$). Note that we assume that the signal from bone only exists in the signal model for the UTE image (i.e., $j = 1$) because of the short $T_2^*$ of bone tissue. The signal model in Eq. 2 has in total five unknowns, i.e., $\rho_W$, $\rho_F$, $\rho_B$, $R_2^*$ and $\Delta f_{B0}$, which requires at least five different TEs for estimation. In our current implementation, we acquire MR signals at seven different TEs in every two TRs (i.e., one UTE and six multi-echo Dixon acquisitions for robust fat-water separation[54]) for more robust parameter estimation. The volume fractions of water, fat, and bone compartments in Eq. 1 can be determined by using the estimated proton densities as follows:

$$p_W = \frac{V_W}{V_W + V_F + V_B} = \frac{\frac{\rho_W}{c_W}}{\frac{\rho_W}{c_W} + \frac{\rho_F}{c_F} + \frac{\rho_B}{c_B}}$$ [3]

$$p_F = \frac{V_F}{V_W + V_F + V_B} = \frac{\frac{\rho_F}{c_F}}{\frac{\rho_W}{c_W} + \frac{\rho_F}{c_F} + \frac{\rho_B}{c_B}}$$ [4]

$$p_B = \frac{V_B}{V_W + V_F + V_B} = \frac{\frac{\rho_B}{c_B}}{\frac{\rho_W}{c_W} + \frac{\rho_F}{c_F} + \frac{\rho_B}{c_B}}$$ [5]

where $V_W$, $V_F$, and $V_B$ each denote the volume occupied by water, fat, and bone within a voxel, respectively, and $c_W$, $c_F$, and $c_B$ each denote the proton concentration of water, fat, and bone, respectively.

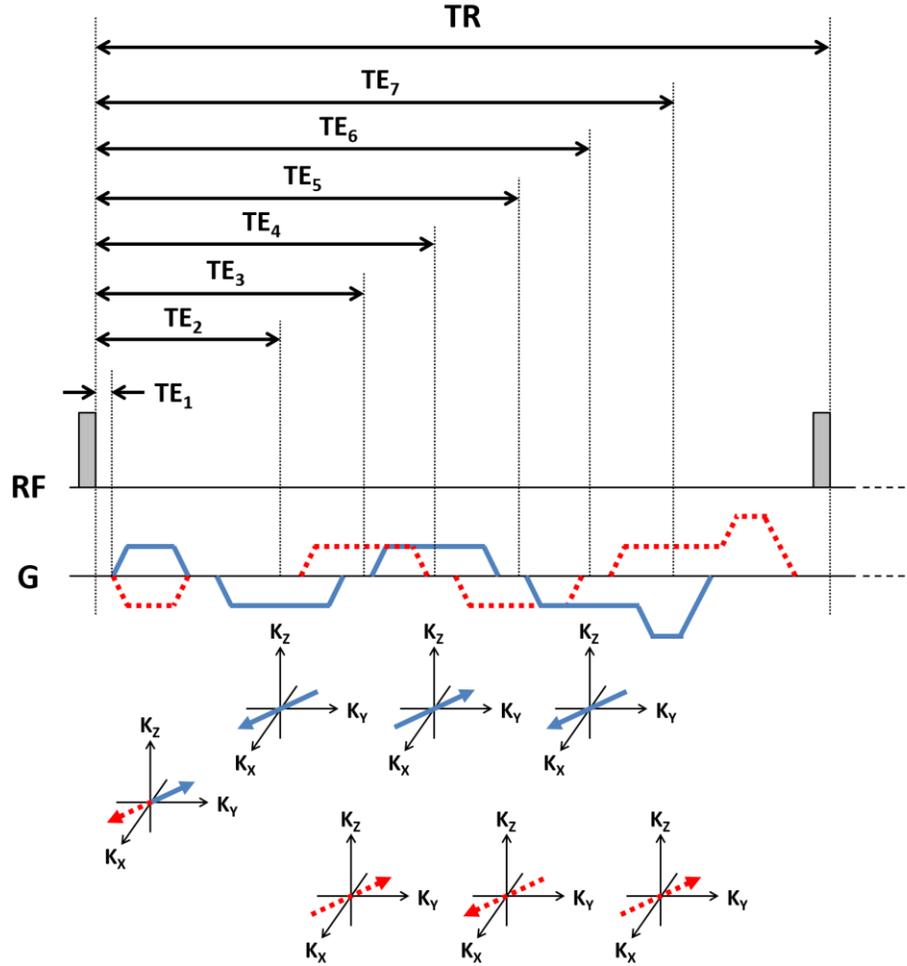

**Figure 1.** Pulse sequence diagram of the proposed mUTE sequence. The proposed mUTE sequence consists of UTE and multi-echo radial images, with alternating acquisition of multi-echo images at different TE times every other TR. Acquisitions from odd TR (e.g., TE$_1$, TE$_3$, TE$_5$, TE$_7$, labeled in red dashed line) and even TR (e.g., TE$_1$, TE$_2$, TE$_4$, TE$_6$, labeled in blue solid line) are shown with a schematic diagram representing the k-space trajectory traversed by each readout gradients. Notice that this interleaved acquisition scheme allows the same k-space coverage of spokes for all images.

*Phantom Experiment*

A phantom was made by placing a pig tibia containing muscle, fat, and bone inside a plastic container. The container was then filled with 5% gelatin mixture containing radioactive $^{18}$F to resemble soft tissue, along with two plastic spheres containing different concentrations of radioactive $^{18}$F to help visualize the location of pig tibia. The phantom was then imaged using a whole-body PET/MR scanner (Biograph mMR, software version VB20P, Siemens Healthcare, Erlangen, Germany). PET imaging was performed for 15 min. MR acquisition was performed using the proposed 3D mUTE sequence (Fig. 1) with the following imaging parameters: field-of-view (FOV) = 240×240×240 mm$^3$, resolution = 1.875×1.875×1.875 mm$^3$, TR = 9.9 ms, TE$_{1-7}$ = 70, 2110, 2810, 3550, 4250, 4990, 5690 μs, flip angle = 10°, hard pulse duration = 100 μs, gradient ramp time = 400 μs (gradient slew rate = 48.9 mT/m/ms), plateau gradient amplitude = 19.57 mT/m, dwell time = 2.5 μs, number of radial spokes for UTE/multi-echo images = 51472/25736, and acquisition time = 8.5 min. Retrospective down-sampling was performed to test the feasibility of the proposed method in the case of radial spokes for UTE/multi-echo images = 6434/3217 (corresponding to an acceleration factor of 8). Additional acquisitions were performed with magnetization-prepared rapid acquisition with gradient echo (MPRAGE) and the manufacturer's two-point Dixon-based MRAC sequence. The phantom was separately imaged using a whole-body CT scanner (Biograph 64, Siemens Healthcare, Erlangen, Germany), with the following parameters: tube peak voltage = 120 kVp, tube current time product = 30 mAs, in-plane resolution = 0.56×0.56 mm$^2$, and slice-thickness = 1 mm.

*In Vivo Experiment*

Six volunteers (1 patient with mild-cognitive impairment and 5 cognitively healthy volunteers; 3 males and 3 females; 48-86 years old) were scanned under a study protocol approved by our local Institutional Review Board (IRB) for Harvard Aging Brain Study (HABS), which includes separate PET/CT and MR scans. Written informed consent was obtained from all subjects before study participation. MR acquisitions were performed on a 3T MR scanner (MAGNETOM Trio, Siemens Healthcare, Erlangen, Germany) using the proposed 3D mUTE sequence (Fig. 1) with the following imaging parameters: FOV = 240×240×240 mm$^3$, resolution = 1.875×1.875×1.875 mm$^3$, TR = 8.0 ms, TE$_{1-7}$ = 70, 2110, 2310, 3550, 3750, 4990, 5190 μs, flip angle = 15°, hard pulse

duration = 100 µs, gradient ramp time = 400 µs (gradient slew rate = 48.9 mT/m/ms), plateau gradient amplitude = 19.57 mT/m, dwell time = 2.5 µs, number of radial spokes for UTE/multi-echo images = 6434/3217 (corresponding to acceleration factor of 8), and acquisition time = 52 s. An additional acquisition was performed with MPRAGE for comparison.

A separate PET/CT examination was performed for the same subjects on a whole-body PET/CT scanner (Discovery MI, GE Healthcare, Milwaukee, Wisconsin, USA). The imaging protocol of PET consisted of 15 mCi bolus injection of [11]C-PiB followed by a dynamic scan for 70 min. Only data from 40-50 min. post injection (i.e., secular equilibrium of [11]C-PiB) were evaluated in the current work. CT imaging was performed with tube peak voltage = 120 kVp, tube current time product = 30 mAs, in-plane resolution = 0.56×0.56 mm$^2$, and slice-thickness = 1 mm.

## Image Reconstruction and LAC Map Generation

In the current work, sparse-sampling was incorporated to accelerate data acquisition and make the total scan time of the proposed mUTE sequence feasible in less than a minute. Image reconstruction based on non-uniform fast Fourier transform (NUFFT)[55] and SENSE[56,57] was performed for UTE and multi-echo radial images by solving the following optimization problem:

$$arg\ min_{x}\|\Omega(\mathcal{F}Sx) - y\|_2^2 + \lambda\|\Psi(x)\|_1 \qquad [6]$$

where the first term denotes data consistency, the second term denotes $l_1$-regularization penalty, $x$ denotes the artifact-free image, $\Omega$ denotes the sampling matrix, $\mathcal{F}$ denotes the spatial Fourier transform matrix, $S$ denotes the coil sensitivity matrix, $y$ denotes the under-sampled k-space measurement, $\lambda$ denotes the regularization parameter, and $\Psi$ denotes the sparsifying transform. Total variation[58] was used for $\Psi$ and the optimization problem was solved using alternating minimization. The regularization parameter $\lambda$ was chosen to be 10$^{-3}$ of the maximum intensity of the original data, based on the data discrepancy principle[59]. The k-space trajectories of the UTE and multi-echo radial acquisitions were measured and corrected in image reconstruction to reduce the effects of eddy currents on the resultant images (See Appendix for details).

Following image reconstruction, variable projection (VARPRO) algorithm[60] was used to estimate $\rho_W$, $\rho_F$, and $\rho_B$ using Eq. 2. Subsequently, $p_W$, $p_F$, and $p_B$ were estimated using Eqs. 3-5, and the LAC was determined for each voxel using Eq. 1. The regions of air were determined via thresholding the UTE image, with an empirically determined threshold value: briefly, the UTE image was bias corrected[61] and an air mask was generated via thresholding with a value determined by evaluating the signal intensity distribution of air[14,32]. For the in vivo studies, an additional air support mask was generated by evaluating the frontal sinus region separately for improved delineation of air, similar to previous study[37]. An additional bone support mask was also generated and applied for improved delineation of bone, via thresholding operation from a bone-enhanced image[10]. An LAC value of 0 cm$^{-1}$ was assigned to voxels determined as air. The values $\mu_W$ of 0.099 cm$^{-1}$ at 511 keV[62] and $c_W$ of 110.4 M[63] were used in this study. Calibration of $\mu_F$, $\mu_B$, $c_F$ and $c_B$ was done by using CT-based attenuation maps at 511 keV from a representative in vivo subject dataset as the ground truth.

For the phantom study, an additional LAC map was generated from the manufacturer's two-point Dixon-based method for comparison. Then, PET images were reconstructed with attenuation correction using the LAC maps. The CT image was converted to LAC at 511 keV using bilinear transformation[64] and was registered to the PET imaging space via registration to MR image first and subsequently to PET using affine transformation (without shearing), prior to reconstruction. The ordered subset expectation maximization (OSEM) algorithm[65] was used with 3 iterations of 21 subsets to reconstruct PET images with the following parameters: image size = 344×344×127, in-plane voxel size = 2.09×2.09 mm$^2$, and slice-thickness = 2.03 mm.

For the in vivo study, additional LAC maps were generated using the two-point Dixon-based method[8], an atlas-based method[17], and the proposed mUTE method with a single LAC assignment of bone for comparison. The multi-echo images from the acquired mUTE sequence at TE$_3$ = 2310 µs and TE$_4$ = 3550 µs were used as in-phase and out-of-phase images for the two-point Dixon-based method, respectively. The same images were used to generate LAC maps using the atlas-based method[17]. The CT image was converted to the LAC at 511 keV using bilinear transformation[64]. Then, PET images were reconstructed with attenuation correction using the LAC maps. All LAC images generated from the MRAC methods were registered to the CT imaging space and then subsequently converted to the PET imaging space prior to reconstruction by using non-rigid registration (with scaling transformation) based on mutual information. The ordered subset expectation maximization (OSEM) algorithm[65] was used with 3 iterations of 17 subsets to reconstruct PET images with the following parameters: image size = 128×128×89, in-plane voxel size = 2.34×2.34 mm$^2$, and slice-thickness = 2.8 mm.

## Analysis

The performance of the proposed mUTE-based MRAC method was evaluated and compared to other MRAC methods, using CT as the ground truth. To evaluate the accuracy of bone LAC estimation, Dice coefficient was calculated for bone using the following equation:

$$Dice\ Coefficient = \frac{2 \cdot |Bone_{MRAC} \cap Bone_{CT}|}{|Bone_{CT}|} \qquad [7]$$

where $Bone_{MRAC}$ and $Bone_{CT}$ each denote the bone region defined from the LAC maps from the MRAC methods and CT, respectively. For the in vivo experiment, structural similarity (SSIM) index[66] was also calculated to assess the similarity of bone LACs between CT and the different MRAC methods that estimate bone LAC. The regions with LAC value above 0.136 cm$^{-1}$ were identified as bone for this analysis[67]. Joint histograms comparing the PET activity after attenuation correction were generated to evaluate the correlation between the PET images reconstructed from CT and the different MRAC methods. Pearson correlation coefficient (r) was calculated and used to determine the coefficient of determination (r$^2$). SSIM index[66] was also calculated to assess the similarity between the PET images reconstructed from CT and the different MRAC methods.

For the in vivo experiment, region-of-interest (ROI) analysis was also performed to evaluate the performance of the proposed method across subjects. Regions of four cortex lobes (i.e., frontal, temporal, parietal, occipital), cerebellum, whole white matter (WM) and whole gray matter (GM) determined by the automated anatomical labeling (AAL) template[68] were evaluated for the analysis. Non-rigid registration (with scaling transformation) based on mutual information was performed to convert the ROIs in the common Montreal Neurological Institute (MNI) space[69] to the imaging space of each individual subject data. Subsequently, the relative absolute error was calculated for each of the ROI and the different MRAC methods in each individual subject as follows:

$$Relative\ Absolute\ Error = \frac{|PET_{MRAC} - PET_{CT}|}{PET_{CT}} \times 100\% \qquad [8]$$

The mean relative absolute error from each ROI was used to calculate the mean and standard deviation across all subjects for the different MRAC methods. Wilcoxon signed rank test was performed to test for significant differences in the performance between the proposed method and the other MRAC methods at significance level of 0.05.

## Results

Figure 2, Supplementary Figures S1 and S2 show the results of k-space trajectory correction using the proposed mUTE sequence. Clear differences were observed between the nominal and measured k-space trajectory traversed by the readout gradients in the mUTE sequence (Supplementary Figure S1c). Noticeable differences were also observed between different gradient axes (Supplementary Figure S1c). With k-space trajectory correction, artifacts were noticeably reduced in all of the images produced by the mUTE sequence (Fig. 2 and Supplementary Figure S2).

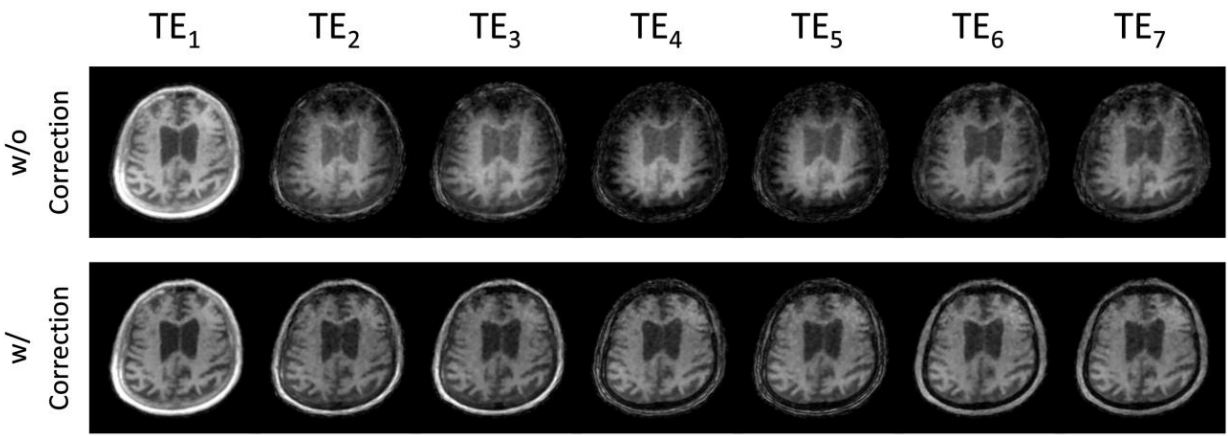

**Figure 2.** Results of k-space trajectory correction for mUTE sequence. Images of mUTE sequence from the in vivo experiment with and without k-space trajectory correction are shown. Notice the difference in UTE ($TE_1$) and multi-echo ($TE_2$, …, $TE_7$) images with and without the k-space trajectory correction.

Figures 3-6 show the results from the phantom experiment. Water, fat, and bone proton density fraction maps were successfully obtained in the expected regions using the proposed mUTE method (Fig. 3). The fraction maps of water and fat obtained from the mUTE method were visually similar to those obtained from the two-point Dixon method, with better separation of water and fat in the gelatin region (Fig. 3). Visually, the LAC map generated from the mUTE method was similar to those from CT, capturing the continuous variation well especially within the bone regions (Fig. 4). The Dice coefficient of bone from the mUTE method was 0.76. The PET images reconstructed from the mUTE method were also similar to those from CT (Fig. 5a), showing lower error compared to those from the two-point Dixon-based method (Fig. 5b). Similar findings were observed in the joint histogram results, with the PET activity reconstructed from the mUTE method showing the best correlation in relation to those reconstructed from CT (Fig. 6). The coefficients of determination ($r^2$) between the reconstructed PET images from CT and the different MRAC methods were 0.97 and 0.99 for Dixon-based method and mUTE method, respectively. The SSIM index between the reconstructed PET images from CT and the different MRAC methods were 0.92 and 0.98 for Dixon-based method and mUTE method, respectively.

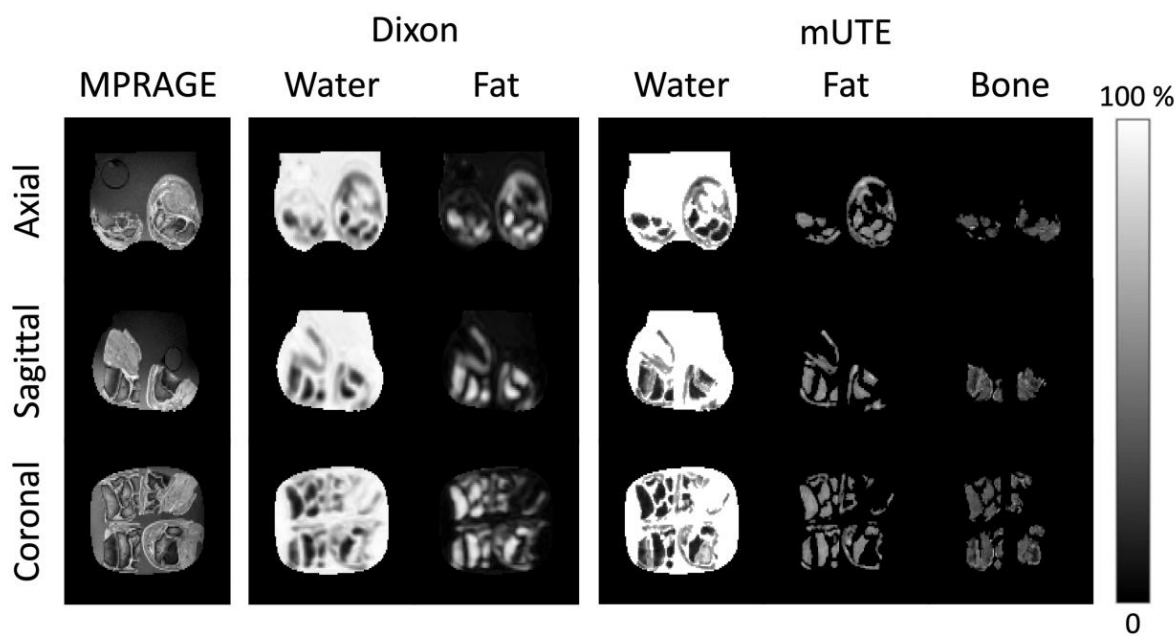

**Figure 3.** Results of quantification from the phantom experiment. MPRAGE images, water and fat proton density fraction maps from the two-point Dixon method, and water, fat, and bone proton density fraction maps from the proposed mUTE method are shown.

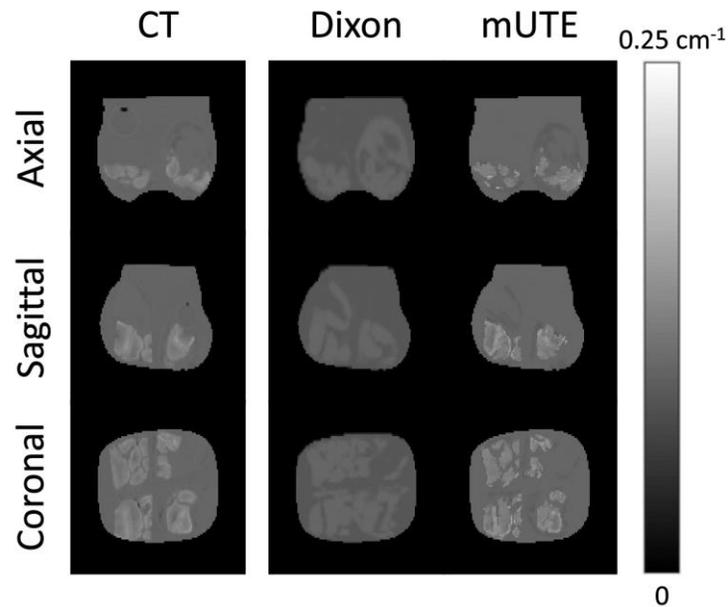

**Figure 4.** Results of LAC maps from the phantom experiment. LAC maps derived from CT, two-point Dixon-based method, and the proposed mUTE method are shown.

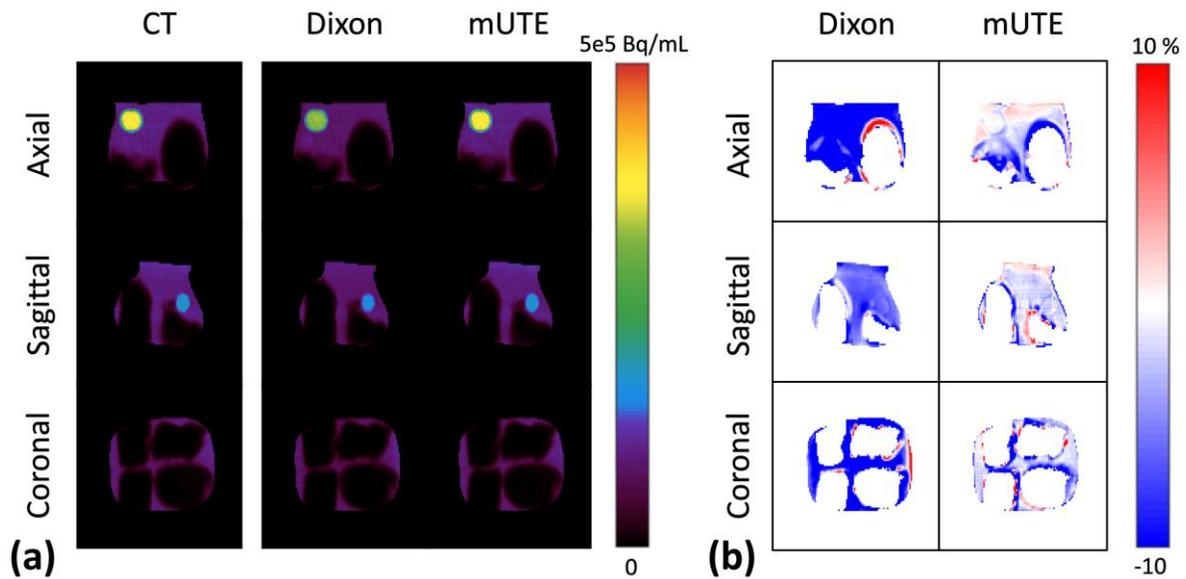

**Figure 5.** Results of PET reconstruction and percentage error maps from the phantom experiment. **a:** PET images reconstructed using LAC maps derived from CT, two-point Dixon-based method, and the proposed mUTE method. **b:** PET percentage error maps for two-point Dixon-based and the proposed mUTE method.

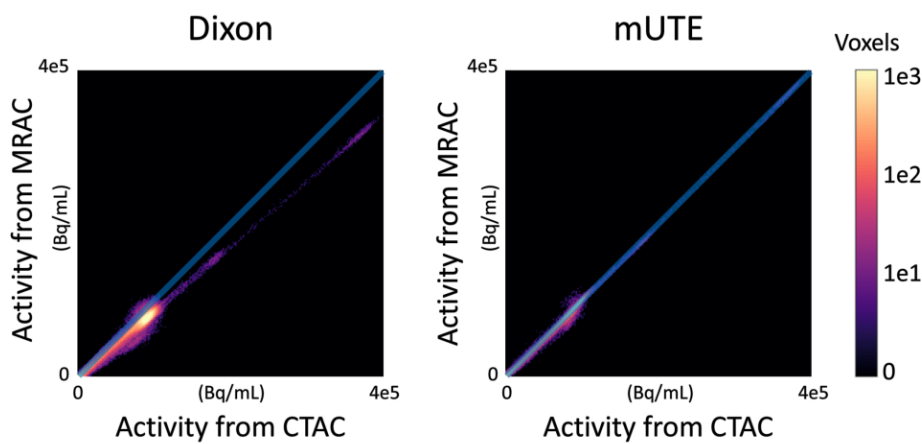

**Figure 6.** Joint histogram analysis results from the phantom experiment. Logarithmic plots of the joint histogram comparing PET activity after attenuation correction between CT and two-point Dixon-based method, and CT and proposed mUTE method are shown. The reference solid line (blue) represents perfect correlation.

In vivo experiment results from a representative subject (Subject #1) are shown in Figures 7-11 and group analysis are shown in Figure 12 and Table 1. Similar to the phantom study results, water, fat, and bone proton density fraction maps were successfully obtained in the expected regions using the proposed mUTE method (Fig. 7 and Supplementary Figure S3). The fraction maps of water and fat obtained from the mUTE method were visually similar to those obtained from the two-point Dixon method, with better separation of water and fat in the brain tissue region (Fig. 7). The LAC map generated from the mUTE method with continuous LAC assignment of bone was visually most similar to those from CT, with better identification of bone/air regions and continuous variation in LAC compared to that from the atlas-based method (Fig. 8). Table 1 summarizes the Dice coefficient and SSIM index of bone LAC for each subjects along with the group results. As can be seen, the mUTE method with continuous LAC assignment of bone achieves higher Dice coefficient and SSIM index than all the other compared methods.

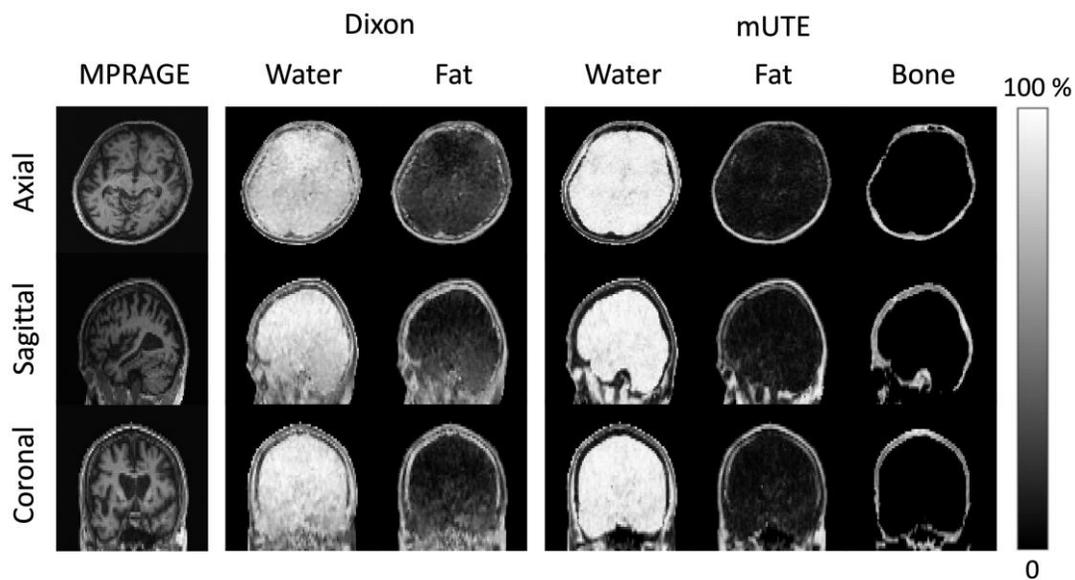

**Figure 7.** Results of quantification from the in vivo experiment (Subject #1). MPRAGE images, water and fat proton density fraction maps from the two-point Dixon method, and water, fat, and bone proton density fraction maps from the proposed mUTE method are shown.

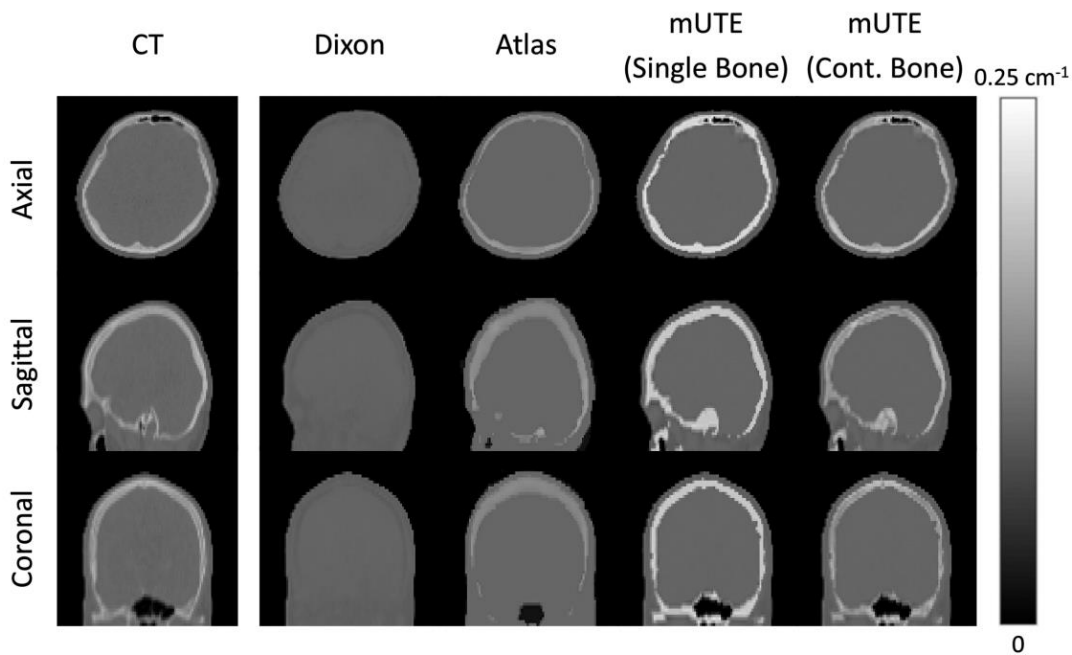

**Figure 8.** Results of LAC maps from the in vivo experiment (Subject #1). LAC maps derived from CT, two-point Dixon-based method, atlas-based method, proposed mUTE method with single LAC assignment of bone, and proposed mUTE method with continuous LAC assignment of bone are shown.

**Table 1.** Quantitative Analysis of AC Map from In Vivo Experiment

| | Dice Coefficient of Bone | | | Structural Similarity (SSIM) Index of Bone | | |
|---|---|---|---|---|---|---|
| | Atlas | mUTE (Single Bone) | mUTE (Cont. Bone) | Atlas | mUTE (Single Bone) | mUTE (Cont. Bone) |
| Subject #1 | 0.56 | 0.75 | 0.77 | 0.74 | 0.70 | 0.82 |
| Subject #2 | 0.56 | 0.76 | 0.78 | 0.66 | 0.65 | 0.80 |
| Subject #3 | 0.42 | 0.66 | 0.69 | 0.66 | 0.64 | 0.78 |
| Subject #4 | 0.59 | 0.88 | 0.90 | 0.77 | 0.77 | 0.90 |
| Subject #5 | 0.47 | 0.76 | 0.77 | 0.69 | 0.67 | 0.82 |
| Subject #6 | 0.36 | 0.88 | 0.90 | 0.72 | 0.80 | 0.88 |
| Group | 0.50 ± 0.09 | 0.78 ± 0.08 | 0.80 ± 0.07 | 0.71 ± 0.04 | 0.70 ± 0.06 | 0.83 ± 0.04 |

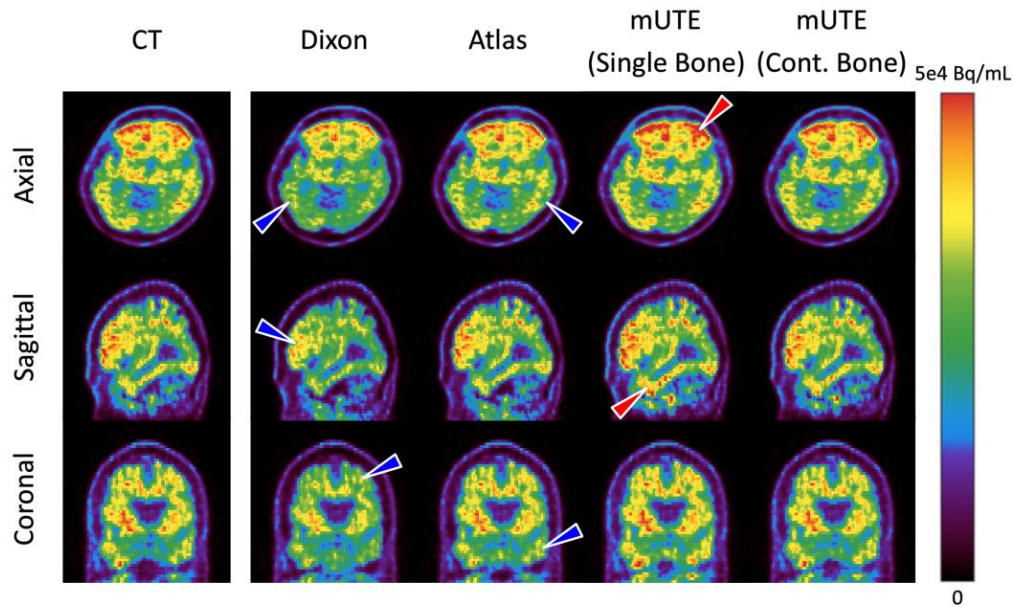

**Figure 9.** Results of PET reconstruction from the in vivo experiment (Subject #1). PET images reconstructed using LAC maps derived from CT, two-point Dixon-based method, atlas-based method, proposed mUTE method with single LAC assignment of bone, and proposed mUTE method with continuous LAC assignment of bone are shown. Notice regions of underestimated PET activity using Dixon- and atlas-based methods (blue arrowhead) and overestimated PET activity using mUTE method with single LAC assignment of bone (red arrowhead).

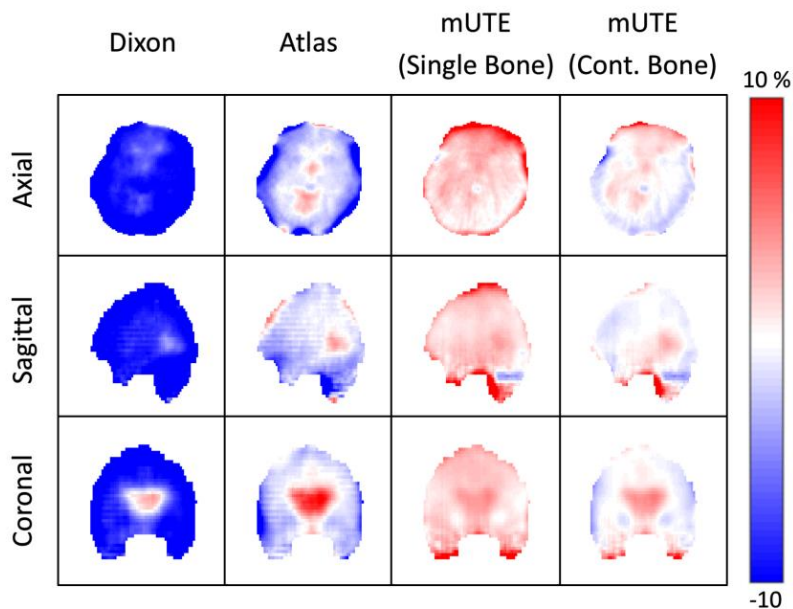

**Figure 10.** Results of PET percentage error maps from the in vivo experiment (Subject #1). PET percentage error maps from two-point Dixon-based method, atlas-based method, proposed mUTE method with single LAC assignment for bone, and proposed mUTE method with continuous LAC assignment for bone are shown.

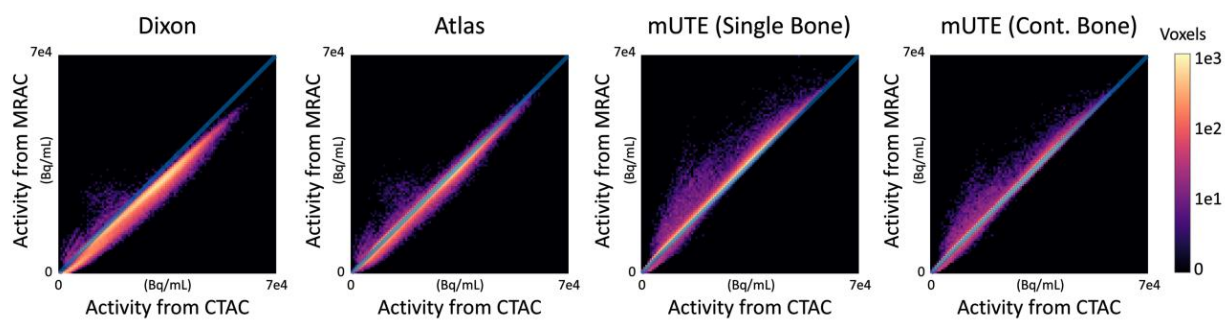

**Figure 11.** Joint histogram analysis results from the in vivo experiment (Subject #1). Logarithmic plots of the joint histogram comparing PET activity after attenuation correction between CT and the two-point Dixon-based method, CT and the atlas-based method, CT and the proposed mUTE method with single LAC assignment of bone, and CT and the proposed mUTE method with continuous LAC assignment of bone. The reference solid line (blue) represents perfect correlation.

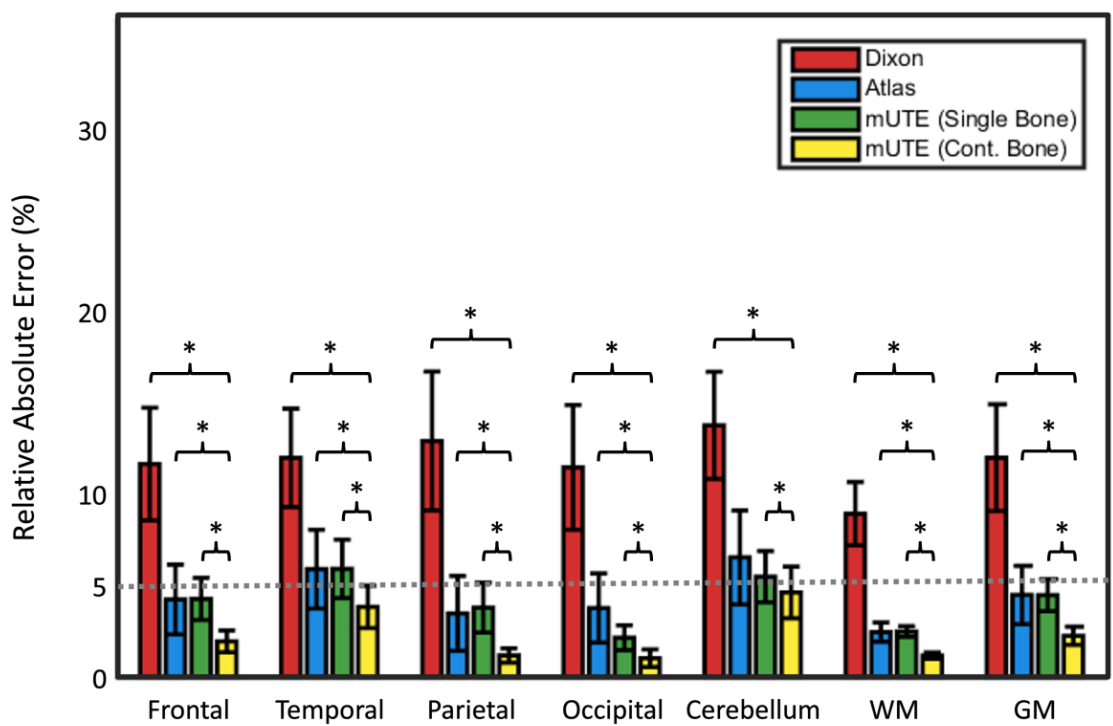

**Figure 12.** ROI analysis result of relative absolute error across all subjects. Relative absolute error calculated at different ROIs across all subjects (n=6) for the two-point Dixon-based method (red), atlas-based method (blue), proposed mUTE method with single LAC assignment of bone (green), and proposed mUTE method with continuous LAC assignment of bone (yellow) are shown along with a reference line at relative error of 5% (gray dashed line). Note that results from using CT for attenuation correction were used as the ground truth for the calculation of relative absolute error. The asterisks denote a statistically significant difference between the proposed method and the other MRAC method at significance level of 0.05.

The PET images reconstructed from the mUTE method with continuous LAC assignment of bone were also similar to those from CT (Fig. 9), showing an overall lowest error across regions compared to those from other MRAC methods (Fig. 10). The PET images reconstructed from the Dixon-based and atlas-based methods showed relatively high underestimation of PET activity in the cortex regions (blue arrowheads in Fig. 9), whereas those from the mUTE method with single LAC assignment of bone showed relatively high overestimation of PET activity in the cortex regions (red arrowheads in Fig. 9). Supplementary Table 1 summarizes the regional mean relative absolute error of the reconstructed PET images from Subject #1 along with the group results. Similar trends were observed in the joint histogram results, with the PET activity reconstructed from the mUTE method showing higher correlation with those reconstructed from CT compared to other methods (Fig. 11). The coefficient of determination ($r^2$) between the reconstructed PET images from CT and the different MRAC methods were 0.95±0.02, 0.96±0.02, 0.95±0.01, and 0.97±0.01 for Dixon-based method, atlas-based method, mUTE method with single LAC assignment of bone, and mUTE method with continuous LAC assignment of bone, respectively. The SSIM index between the reconstructed PET images from CT and the different MRAC methods were 0.95±0.02, 0.98±0.01, 0.98±0.01, and 0.99±0.01 for the Dixon-based method, atlas-based method, mUTE method with single LAC assignment of bone, and mUTE method with continuous LAC assignment of bone, respectively. The ROI analysis (Fig. 12) showed the best performance for the mUTE method with continuous LAC assignment of bone, resulting in mean relative absolute error below 5% with the lowest standard deviation compared to other methods in all of the ROIs. Statistically significant differences were observed between the proposed method and the other MRAC methods at all ROIs, except the cerebellum which did not show a statistically significant difference between the proposed method and the atlas-based method.

## Discussion

In this work, a physical model-based MRAC method is proposed for PET/MR. UTE and multi-echo Dixon images are used with a physical compartmental model to estimate the fractions of water, fat and bone components, which are subsequently used to estimate the continuous LAC maps for PET attenuation correction. Although the combination of UTE/ZTE with Dixon has been proposed in the past for PET attenuation correction[13,14,33,70], the current work differs from previous approaches in the sense that (i) a new sequence integrating 3D UTE with multi-echo Dixon imaging in a single acquisition is developed, (ii) a general physical compartmental model is used to estimate the signal from the different compartments of water, fat and bone altogether, and (iii) continuous variation of LAC is estimated via the different proportions of compartments. The proposed method does not require prior anatomical structure information for LAC estimation and is robust to $B_0$ and receive $B_1$ effects.

The proposed method showed better performance in comparison to the conventional Dixon-based and atlas-based MRAC methods, in both the phantom and in vivo human studies. The proposed method not only reconstructed PET images with higher correlation and similarity, but also with lower mean and standard deviation of relative absolute error across different ROIs and subjects compared to other MRAC methods (Fig. 12). Since the choice of LACs for different compartments may influence the magnitude of relative absolute error but not the variation, assessing the variation of error may be more important in evaluating the accuracy of PET attenuation correction methods. Although tested from a small number of subjects (n=6), the current study showed the smallest variation of relative absolute error using the proposed method (Fig. 12), indicating better PET attenuation correction. However, the proposed method did not show statistically significant difference compared to the atlas-based MRAC method for the cerebellum, which is presumed to be due to the small number of subjects (n=6). Further investigations are necessary in increased number of subjects to confirm the findings from this work regarding the performance of the proposed method.

Trajectory correction is necessary to accurately reconstruct the UTE and multi-echo Dixon images and estimate the different compartments using the proposed method. Since the k-space trajectory traversed by the readout gradients of mUTE can be different for each gradient axis due to the distortion by hardware imperfections and eddy current effects (Supplementary Figure S1c), a simple gradient delay correction[71] may not be sufficient for accurate image reconstruction (Supplementary Figure S2). The k-space trajectory can be measured from all three gradient axes to improve the performance of correction (Supplementary Figure S2). Note that, the employed trajectory correction method is intrinsically robust to phase-wrapping by design: phase wrap-around effects can be minimized by determining $\emptyset(t)$ recursively from $\emptyset(\Delta t_N)$, since $\emptyset_G(\Delta t_N) = \emptyset_G(t_N) - \emptyset_G(t_{N-1})$ will always be lower than $\pi$ as long as the sampling frequency satisfies the Nyquist limit for the gradient structure of interest. It must be noted that the number of phase-encodings ($N_{PE}$) in the employed trajectory correction sequence must be chosen carefully to satisfy $N_{PE} \geq \max\{k_G(t)\} \cdot \text{FOV}$, to properly resolve a phase shift of $\pi$ produced over each voxel at the maximum value of $k_G(t)$ [72].

The current work has several limitations. The effects of $T_1$ and flip angle were neglected in the proposed signal model for parameter estimation, which may not be negligible when the flip angle is not small enough or TR is not short enough as in the current study. However, this error may not be a problem for PET attenuation correction, since the ratio between the estimated parameters is evaluated for the generation of LAC maps. This error can be mitigated by choosing a smaller flip angle and shorter TR for acquisition, or by acquiring a separate acquisition with $T_1$ mapping. Another source of error exists in the estimation of the short-$T_2$ component. In the current work, the $R_2^*$ effect due to short-$T_2$ component was neglected and the proton density of bone was estimated relying on a single UTE acquisition. This issue can be potentially resolved by modifying the sequence to acquire a larger number of UTE images at different TE times and taking into account the $R_2^*$ effect due to the short-$T_2$ component. Also, in the current work the MR scans for the in vivo experiments were conducted using an MR scanner instead of a PET/MR scanner, due to the clinical workflow and associated study protocol. However, we don't expect any major difference between the mUTE images obtained from the MR scanner and the PET/MR scanner, since the mUTE sequence used in the in vivo experiments on the MR scanner was implemented with the same hardware constraints (i.e., minimum TE, maximum gradient strengths and maximum gradient slew rates) as the mUTE sequence used in the phantom experiment on the PET/MR scanner. We have also carried out phantom experiments to compare the signal-to-noise ratio (SNR) of the images acquired on the two scanners using the same phantom and same sequence and found very comparable SNRs (results not shown). Nevertheless, further investigation with in vivo experiments may be necessary to further validate the performance of the proposed method on PET/MR scanners. Lastly, in the current work CT-based attenuation correction map was used as the ground truth reference for comparison, which may contain systematic bias that can result in overestimation of the PET activity[73]. Further comparisons with attenuation maps from 511 keV transmission scans may be necessary to accurately characterize the performance of the proposed method.

Several aspects of the proposed method can be further improved. In this work, the TE times and the number of echoes were chosen arbitrarily for UTE and the multi-echo Dixon images (i.e., one UTE and six multi-echo Dixon acquisitions for robust fat-water separation). Cramér–Rao bound (CRB) analysis can be performed to determine the optimal TE times/number of echoes and improve the estimation performance of the proposed method. The acquisition time of the proposed method may also be further reduced through optimizations. For example, for the case of the proposed physical model where five parameters need to be estimated, signal from five echoes (including UTE) are needed at minimum which can reduce the scan time further (e.g., scan time reduction by 25%). A preliminary study was performed to demonstrate the feasibility of the proposed method using only five echoes (Supplementary Figure S4), which requires further investigation for validation. More advanced image reconstruction with higher under-sampling factor can reduce the scan time further. Also, a voxel-based least squares approach is used in the current work to estimate the parameters. Parameter estimation can be further improved with enhanced robustness to $B_0$ inhomogeneity using joint estimation with constraints[74], with and without region-growing[75] or graph-cut algorithms[76]. Moreover, the proposed approach is not mutually exclusive to other methods and can potentially be improved by combining with other approaches. In this work, segmentation was used as a preprocessing step to improve the delineation of bone and air, i.e., by using bone-enhanced image for bone[10] and applying a differential thresholding strategy to the frontal sinus region for air[34]. Similarly, the proposed approach can be further improved by additionally utilizing information from bone-enhanced images generated from different strategies[11,13], and by incorporating a more rigorous region-based differential thresholding strategy for the delineation of air, especially for regions such as the ethmoidal sinuses, nasal septa and mastoid process[37]. Also, combination with DNN-based algorithms may further improve the performance of the proposed method. Recently, DNN-based algorithms utilizing information from both ZTE and two-point Dixon have been developed and showed outstanding performance for MR-based PET attenuation correction in the brain[47] and pelvis[44,45] regions. Since the proposed method allows acquisition of multiple images of UTE and multi-echo Dixon along with additional information such as the proton densities of different compartments as well as $B_0$ and $R_2^*$ via parameter estimation (Supplementary Figure S3), the proposed method has great potential to work well with DNN-based algorithms.

The proposed method may be extended for application to other parts of the body. However, several technical challenges may arise. First, the greater variety of tissue types need to be considered. For instance, trabecular bone in the body consists of more water and fat components compared to cortical bone (e.g., bone in the skull) due to the porous structure filled with bone marrow and blood vessels. More complicated compartmental model may be needed for the proposed method to handle trabecular bone and/or the different tissue types. Second, respiratory motion may be a concern. This issue could be addressed by further accelerating the imaging speed and by the usage of breath-hold. Third, $B_0$ inhomogeneity in other body parts may be more severe than the brain and could be a concern. The proposed method uses multi-echo Dixon method for water/fat separation and is relatively robust to $B_0$ inhomogeneity. Further investigation is necessary to verify the feasibility of the proposed method to other body regions.

## Conclusions

A physical compartmental model-based MRAC method was developed for PET/MR. The proposed method generates subject-specific, continuous LAC maps for quantitative PET image reconstruction in PET/MR.

## Acknowledgements

This work was supported in part by the National Institutes of Health (T32EB013180, R01CA165221, R01HL118261, R21EB021710, and P41EB022544) and the Federal Share of Program Income earned by the Massachusetts General Hospital on C06CA059267.

## Conflict of Interest

The authors have no relevant conflicts of interest to disclose.

## Appendix: k-Space Trajectory Correction for mUTE

The k-space trajectories of the UTE and multi-echo radial acquisitions were measured and corrected in image reconstruction to reduce the effects of eddy currents on the resultant images. Our k-space trajectory mapping method is similar to that in [1,2], which measures the phase accrual at each time point throughout various spatial locations to map the actual k-space trajectory traversed by the readout gradient of interest. The pulse sequence diagram of the employed k-space trajectory correction sequence is shown in Supplementary Figure S1. Spatial localization was achieved using a slice-selective RF excitation pulse followed by phase-encoding gradients in two directions for 2D imaging. Free-inductive-decay (FID) signals were subsequently acquired to measure the phase accrual caused by magnetic field inhomogeneity alone ("Gradient-Off") and the additional phase accrual caused by the gradient structure of interest ("Gradient-On"). More specifically, the signal obtained in the "Gradient-On" acquisition at spatial location $x$ and time $t$ of the acquisitions can be written as:

$$S(x,t) = I(x) \cdot e^{-i(\emptyset(t))} \tag{1}$$

The phase term $\emptyset(t)$ consists of two terms:

$$\emptyset(t) = \emptyset_{B_0}(t) + \emptyset_G(t) \tag{2}$$

where $\emptyset_{B_0}(t)$ denotes the phase accrual at time $t$ due to magnetic field inhomogeneity $B_0$ and $\emptyset_G(t)$ denotes the phase accrual at time $t$ due to the execution of the gradient structure of interest. The phase contribution due to $\emptyset_{B_0}(t)$, i.e., $\emptyset_{B_0}(t) = 2\pi \cdot \Delta f_{B0} \cdot t$, can be removed via subtraction of $\emptyset(t)$ between acquisitions with (Supplementary Figure S1a) and without (Supplementary Figure S1b) the execution of the gradient structure of interest. Then, the k-space trajectory traversed by the gradients can be estimated by further writing $\emptyset_G(t)$ as:

$$\emptyset_G(t) = x \cdot \gamma \cdot \int_{t_0}^{t} G(\tau)d\tau = 2\pi \cdot x \cdot k_G(t) \tag{3}$$

where $\gamma$ denotes the gyromagnetic ratio, $t_0$ denotes the starting time of the gradient $G$, and $k_G(t)$ denotes the time evolution of the k-space trajectory traversed by the readout gradients of interest.

The k-space trajectory was measured for the proposed mUTE sequence by performing experiments with a water phantom on the scanners used for phantom and in vivo studies, the whole-body 3T MR scanner (MAGNETOM Trio, Siemens Healthcare, Erlangen, Germany) and whole-body PET/MR scanner (Biograph mMR, Siemens, Erlangen, Germany), using a head coil for reception and body RF coil for transmission. The imaging parameters were: field-of-view (FOV) = 240×240 mm$^2$, resolution = 1.875×1.875 mm$^2$, slice thickness = 5 mm, TR/TE = 14/2.6 ms, and total scan time = 15.3 min. The TR unit with (Supplementary Figure S1a) and without (Supplementary Figure S1b) the readout gradient structure of the mUTE sequence was acquired in an alternating fashion for each of the phase-encoding gradients, and the entire acquisition was repeated for readouts along the $x$-, $y$-, and $z$-gradient axes. A one-time acquisition was performed from each scanner to measure the k-space trajectory for the mUTE sequence, and the measured k-space trajectory was applied for reconstruction of all subsequent acquisitions performed across several months with the mUTE sequence from the same scanner. Following the acquisition, the phase difference between each time point (i.e., $\emptyset(\Delta t_N) = \emptyset(t_N) - \emptyset(t_{N-1})$) was evaluated first with $\emptyset(t_0) = 0$ and was subsequently used to determine $\emptyset(t)$ recursively to minimize effects from phase-wrapping. Note that for the mUTE sequence, data sampling starts from the center of the k-space. The phase contribution due to $\emptyset_{B_0}(t)$ was removed by subtracting $\emptyset(t)$ between acquisitions with (Supplementary Figure S1a) and without (Supplementary Figure S1b) the readout gradients of the mUTE sequence. Subsequently, $k_G(t)$ was derived via linear fitting with least-squares method.

**Supplementary Figures and Table**

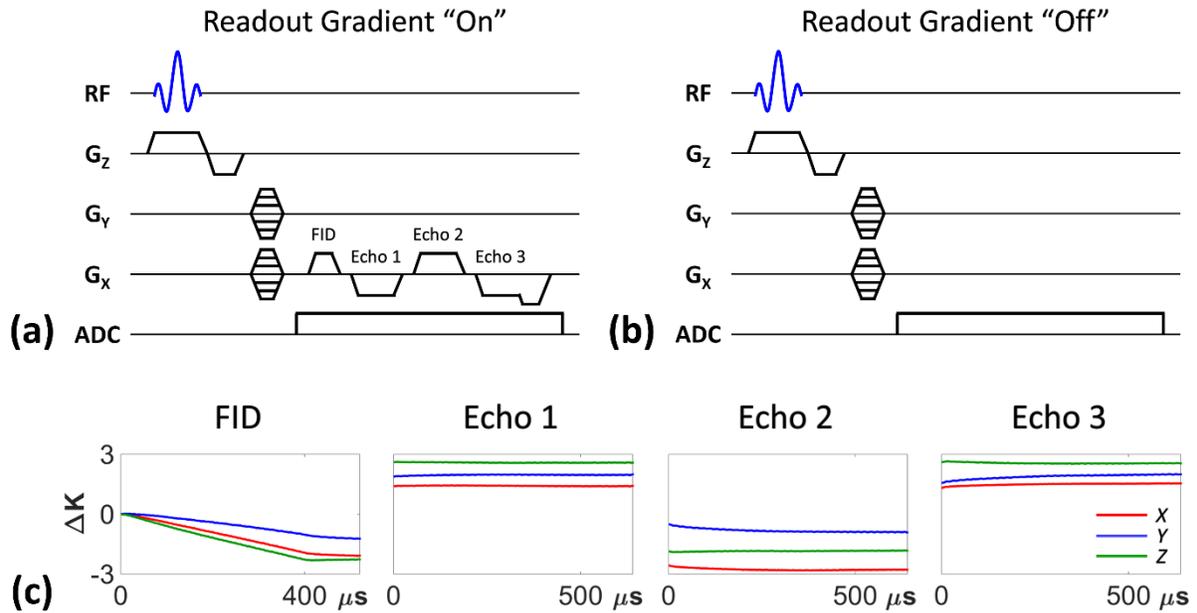

**Supplementary Figure S1.** The k-space trajectory mapping sequence. **a:** Pulse sequence diagram of the k-space trajectory mapping sequence with the gradient structure of interest (e.g., readout gradients in the mUTE sequence). **b:** Pulse sequence diagram of the k-space trajectory mapping sequence without the gradient structure of interest. **c:** Nominal and measured k-space trajectory of the readout gradients in the mUTE sequence from $x$- (red), $y$- (blue) and $z$- (green) gradient axes. Notice the difference in measurement from the different axes.

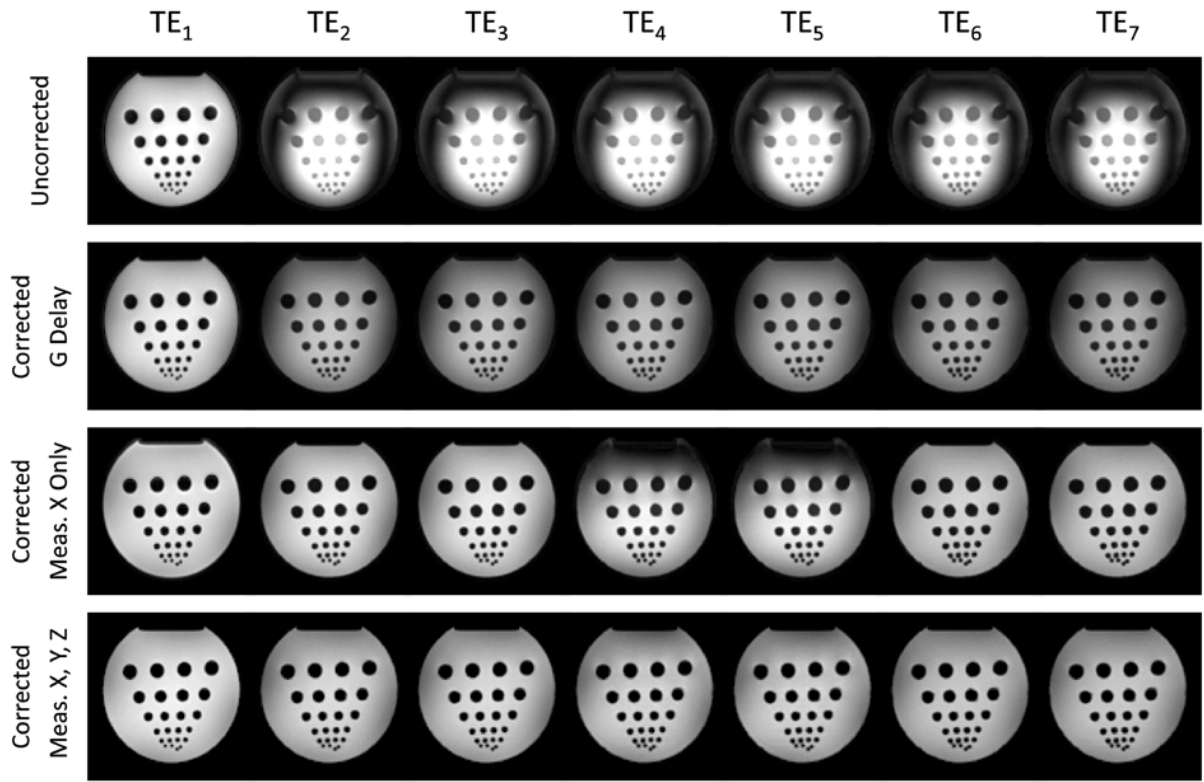

**Supplementary Figure S2.** Phantom experiment results of k-space trajectory correction for mUTE sequence. Images reconstructed without k-space trajectory correction and with k-space trajectory correction using gradient delay, measurements from the *x*-gradient axis only, and measurements from *x*-, *y*-, and *z*-gradient axes are shown. Notice the remaining signal loss and distortion in images when k-space trajectory was corrected using gradient delay or measurements from the *x*-gradient axis only. Also notice the improvement when k-space trajectory was corrected using measurements from *x*-, *y*-, and *z*-gradient axes.

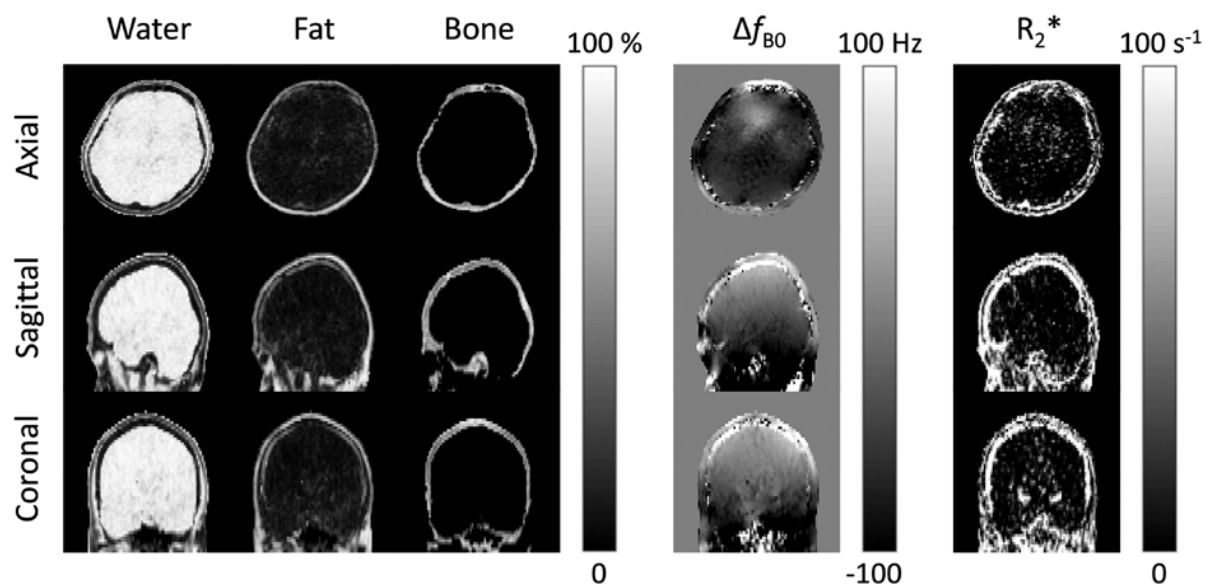

**Supplementary Figure S3.** Estimation results of the proposed mUTE method (Subject #1). Estimated water, fat, and bone proton density fraction, $\Delta f_{B0}$, and $R_2*$ maps from the proposed mUTE method are shown.

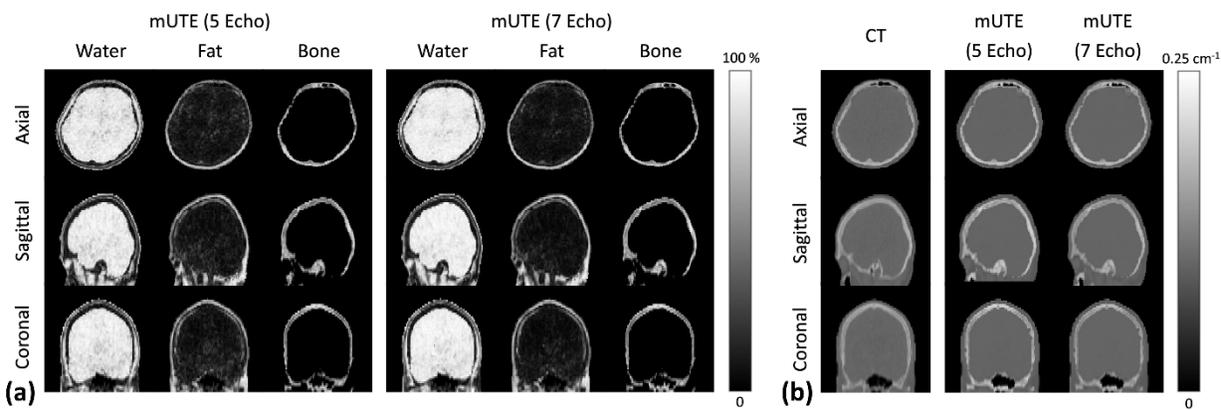

**Supplementary Figure S4.** Results of quantification and LAC maps from the proposed mUTE method with different number of echoes (Subject #1). **a:** Water, fat, and bone proton density fraction maps from the proposed mUTE method using 5 and 7 echoes. **b:** LAC maps from the proposed mUTE method using 5 and 7 echoes.

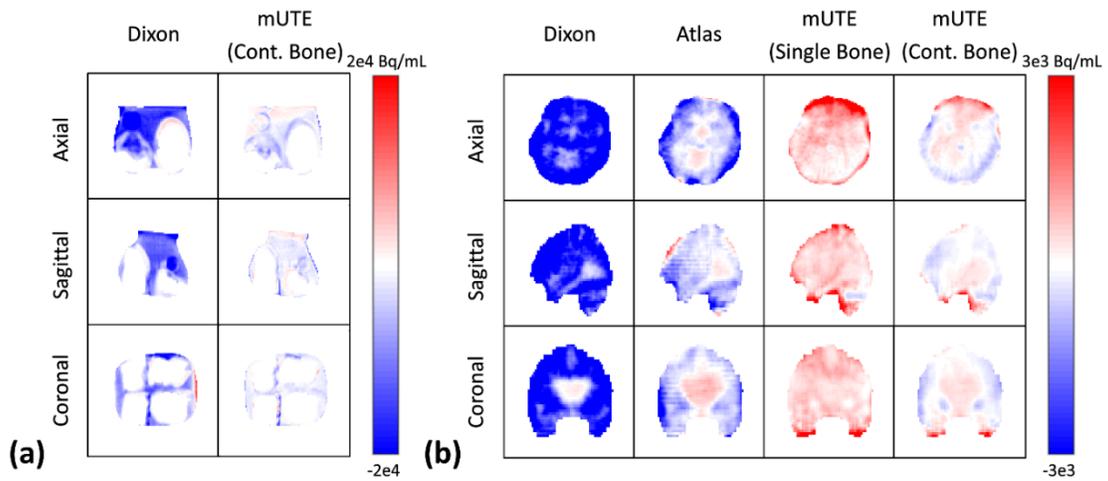

**Supplementary Figure S5.** Results of PET percentage error maps in absolute units. **a**: PET percentage error maps from the phantom experiment. **b**: PET percentage error maps from the in vivo experiment (Subject #1).

**Supplementary Table 1**. Regional Mean Relative Absolute Error of Reconstructed PET Images from Subject #1

|  | Mean Relative Absolute Error (%) | | | |
|---|---|---|---|---|
|  | Dixon | Atlas | mUTE (Single Bone) | mUTE (Cont. Bone) |
| Frontal | 13.7 (11.7 ± 3.1) | 4.8 (4.3 ± 1.9) | 6.1 (4.3 ± 1.2) | 2.9 (2.0 ± 0.6) |
| Temporal | 12.1 (12.0 ± 2.7) | 5.9 (5.9 ± 2.2) | 8.5 (5.9 ± 1.6) | 5.6 (3.9 ± 1.2) |
| Parietal | 15.0 (12.9 ± 3.8) | 3.1 (3.5 ± 2.1) | 3.6 (3.8 ± 1.4) | 1.0 (1.2 ± 0.4) |
| Occipital | 13.0 (11.5 ± 3.4) | 3.4 (3.8 ± 1.9) | 2.2 (2.2 ± 0.7) | 0.9 (1.1 ± 0.5) |
| Cerebellum | 13.9 (13.8 ± 2.9) | 7.9 (6.6 ± 2.6) | 7.1 (5.5 ± 1.4) | 5.5 (4.7 ± 1.4) |
| WM | 8.9 (9.0 ± 1.7) | 2.2 (2.5 ± 0.5) | 2.9 (2.5 ± 0.3) | 1.4 (1.2 ± 0.2) |
| GM | 13.5 (12.0 ± 2.9) | 4.6 (4.5 ± 1.6) | 5.9 (4.5 ± 0.9) | 3.1 (2.3 ± 0.5) |

[a] Group result shown in parenthesis (mean ± standard deviation)